\documentclass[aps,pra,twocolumn,floatfix,amsmath,amssymb,eqsecnum,nofootinbib,superscriptaddress]{revtex4-1}
\usepackage{graphicx}% Include figure files
\usepackage{dcolumn}% Align table columns on decimal point
\usepackage{bm}% bold math
\usepackage{epstopdf}
\usepackage{framed}
\usepackage[toc,page]{appendix}

\usepackage[utf8]{inputenc}

\newcommand{\beq}{\begin{equation}}
\newcommand{\eeq}{\end{equation}}
\newcommand{\ba}{\begin{array}{ccc}}
\newcommand{\ea}{\end{array}}

\def\bea{\begin{eqnarray}}
\def\eea{\end{eqnarray}}

\def\Tr{ {\rm Tr} }
\def\<{\langle}
\def\>{\rangle}

\def\bZ{\mathbb{Z}}
\def\bR{\mathbb{R}}

\usepackage{amsmath}
\usepackage{amssymb}
\usepackage{amsthm}

\theoremstyle{definition}

\usepackage{color}
\usepackage{tikz-cd}
\usepackage{enumerate}
\usepackage[caption=false]{subfig}

\usepackage{tikz}
\usetikzlibrary{decorations.markings}
\begin{document}

\title{Topological Terms and Phases of Sigma Models}
\author{Ryan Thorngren}
\affiliation{UC Berkeley}
\begin{abstract}
We study boundary conditions of topological sigma models with the goal of generalizing the concepts of anomalous symmetry and symmetry protected topological order. We find a version of 't Hooft's anomaly matching conditions on the renormalization group flow of boundaries of invertible topological sigma models and discuss several examples of anomalous boundary theories. We also comment on bulk topological transitions in dynamical sigma models and argue that one can, with care, use topological data to draw sigma model phase diagrams.
\end{abstract}
\maketitle

In recent years, there has been an explosion of activity in both condensed matter \cite{KaneMele,Bernevigetal,SenthilReview} and high energy physics \cite{GKKS,GKS,Seibergetal} surrounding discrete symmetries and their anomalies. In particular, a correspondence has been established between $G$-symmetric systems with unique ground state and mass gap and anomalous $G$ symmetries of systems in one smaller dimension (the ``bulk-boundary correspondence" or more specifically ``anomaly in-flow" generalizing classic ideas when $G$ is connected \cite{CallanHarvey,FaddeevShatashvili}). This correspondence relies on the study of boundary conditions for topological terms in $G$ gauge theories. The purpose of this note is to explore boundary conditions for topological terms in other types of theories, especially sigma models, and to attempt to derive physical constraints on boundaries and phase diagrams of gapless systems.

Indeed, one can consider a $G$ gauge theory as a kind of sigma model, since a $G$ gauge field over a spacetime $M$ is the same thing as a map $A:M \to BG$, where $BG$ is a typically infinite-dimensional space known as the classifying space of $G$. This space is easiest to construct for finite $G$ \cite{Hatcher}, though constructions exist for Lie groups as well and can even encode the configuration space of dynamical gauge theories\cite{Schreiber}. Dijkgraaf and Witten used this approach to study topological terms of these theories by studying the cohomology of $BG$ \cite{DW}. If the spacetime dimension is $D$, then any cohomology class $\omega \in H^D(BG,U(1))$ defines a topological term by pullback:
\[S(A) = {\rm kinetic\ and\ potential\ terms\ } + \int_M A^*\omega.\]
When $M$ is not closed, the resulting term is gauge invariant only up to a boundary variation. Thus, we must either restrict the class of boundary gauge transformations (by breaking the symmetry) or include new boundary degrees of freedom which are invariant under global $G$ transformations but not local ones. In the latter case, the boundary system is said to have an 't Hooft anomaly and we say the combined system is gauge invariant by anomaly in-flow \cite{tHooft}. This typically implies the symmetric boundary is either gapless or carries topological order \cite{CGLW}.

What happens when $BG$ is replaced by some other space $X$? We can study topological terms of such theories using the same techniques, do these lead to non-trivial constraints on the boundary modes? What does anomaly mean in these situations where there is no symmetry at play, only the topology of the configuration space?

\section{phases of sigma models}

We consider field theories of maps $\sigma:M \to X$, where $M$ is the spacetime and $X$ is the target space. Written as a path integral, the (Euclidean) partition $Z(M)$ function decomposes as a sum over homotopy classes of $\sigma$. We denote the \emph{space} of such maps as $Maps(M,X)$ and the set of homotopy classes are the connected components of this space $\pi_0 Maps(M,X)$, with $[\sigma]$ denoting the homotopy class of $\sigma$. Then we may write
\begin{equation}\label{partitionfunction}
Z(M) = \sum_{[\sigma]} Z(M,[\sigma]).
\end{equation}
We will assume that a regularization procedure exists which can define all of the $Z(M,[\sigma])$ separately. In other words, while soliton ``number" $[\sigma]$ is a classically conserved quantity, we need to assume that it remains conserved in the quantum theory.

We are interested in global features of this sum. There are three interrelated approaches, given in increasing difficulty but desirability. The first is to take $Z = Z_{top}$ to be topological, that is, independent of the metric or other geometric data of $M$ up to continuous deformation. This is useful for studying boundary theories, whose RG flows will be constrained by a version of 't Hooft anomaly matching we discuss in Section \ref{invertible}.

The second approach is to take a dynamical sigma model $Z_0$ and form the \emph{twisted sigma model}
\[Z_{twist}(M,[\sigma]) = Z_0(M,[\sigma]) Z_{top}(M,[\sigma])\]
analogous to how one may twist a gauge theory by a Dijkgraaf-Witten term \cite{AST,KapustinSeiberg,Scaffidietal,Pollmannetal}. We then ask how is $Z_{twist}$ different from $Z_0$? For instance, we expect that boundary conditions for $Z_0$ and $Z_{top}$ will combine to form interesting boundary conditions of $Z_{twist}$.

The third approach is to begin with any sigma model and try to factor it as a twisted sigma model, extracting the topological part. This is the situation we must study to derive constraints on the phase diagram of dynamical sigma models.

For example, we will study situations where solitons carry $G$ charges. This is detected by the leading term in a twisted partition function on $M = N \times S^1$, where $N$ is a compact space:
\begin{equation}\label{gap}
Z(N \times S^1, g, [\sigma]) = \Tr_{N,[\sigma]} U(g)\ e^{-\beta H} = \chi(g) e^{-\beta E_0} + \cdots
\end{equation}
where $[\sigma] \in \pi_0 Maps(N,X)$ is the soliton number labelling the sector of Hilbert space of $N$ over which the trace is computed, $U(g)$ is the unitary representation of a symmetry element $g \in G$ on the Hilbert space, $\beta$ is the length of the thermal circle, $\chi(g)$ is a character of $G$, telling us what charge the soliton carries, $E_0$ is the zero-point energy of the soliton, and the dots denote subleading exponentials in $\beta$. Indeed, since $N$ is compact, there is always a gap above the soliton ground states, so $\chi(g)$ can only change at special values of system parameters where this gap closes. We discuss an example of such a topological transition in Section \ref{transition}. One effect of a topological twist is to change what character appears in this expansion.

In what follows, we will consider sigma models which are \emph{nondegenerate}, meaning the ground state is nondegenerate in each topological sector $\sigma \in \pi_0 Maps(N,X)$ and all compact spaces $N$. For twisted sigma models this condition implies that $Z_{top}(M,[\sigma])$ is an \emph{invertible} topological sigma model (for which $Z_{top}(M,[\sigma])^{-1}$ is also the partition function of a topological sigma model). Such theories were recently studied using homotopy theory in a related context in \cite{FKS} and in an attempt to classify crystalline SPT phases in \cite{cspt}.

\section{Invertible topological sigma models}\label{invertible}

The most basic invertible topological sigma models are the analogs of Dijkgraaf-Witten theory \cite{DW}: they are given by a choice of target space $X$ and a cocycle $\omega \in H^m(X,U(1))$, where $m$ is the spacetime dimension. The partition function of a map $\sigma:M \to X$ is a homotopy invariant (provided $M$ is closed), defined by integration of the pullback of $\omega$:
\begin{equation}\label{topresp}
Z_{top}(M,[\sigma]) = \exp\left(i\int_M \sigma^*\omega\right).
\end{equation}
Dijkgraaf-Witten theory is a special case where we take $X = BG$. Note that for $X$ a closed manifold of dimension $n$, nontrivial twisting cocycles $\omega$ only exist if $n>m$.

In analogy with invertible gauge theories \cite{K,KTTW}, we expect the most general invertible topological sigma models are classified by the group $\Omega^m_{str}(X)$ of $U(1)$-valued cobordism invariants of manifolds $M$ with a map $\sigma:M \to X$ and particular tangent structure noted in the subscript. $Z \in \Omega^m_{str}(X)$ iff it is multiplicative over disjoint unions and for all $m+1$ manifolds $W$ with a map $\sigma:W \to X$ and specified tangent structure, $Z(\partial W) = 1$, where the boundaries acquire their maps to $X$ and tangent structures by restriction. In the simplest cases, this tangent structure is an orientation for bosonic systems and a spin structure for fermion systems, though more general twisted tangent structures have been physically relevant in classification of invertible topological gauge theories and SPT phases \cite{KTTW,cspt}.

While all such cobordism invariants $Z$ appear as the partition function of a topological sigma model, some cannot be written as integrals of local densities. These ``Cheshire charges" \cite{WilczekCheshire,ElseCheshire} describe how Skyrmion-like defects couple to the tangent bundle of $M$. For example, the $\pi$ Hopf-term \cite{FKS,PolyakovWiegmann}, which makes sense only in fermionic systems, lives in $\Omega^3_{spin}(S^2) = \bZ_2$. Using the Atiyah-Hirzebruch spectral sequence \cite{GaiottoKapustin,BhardwajGaiottoKapustin,KapustinThorngrenBosonz} its origin can be traced to $H^2(S^2,\Omega^1_{spin}(\star)) = H^2(S^2,\bZ_2) = \bZ_2$ which indicates that the Skyrmion has odd fermion parity. We will return to this and similar examples below.

\subsection{Boundary Variation and Anomaly}

Interesting things begin to happen when we consider $M$ with boundary. First we recall what happens in Dijkgraaf-Witten theory. Let $G$ denote the gauge group and $A$ a $G$ gauge field on $M$ with twisting cocycle $\omega(A) \in H^m(M,U(1))$. In presence of a boundary, the action \eqref{topresp} is not gauge invariant. Instead,
\begin{equation}\label{boundarygaugevariation}
\int_M \omega(A^g) = \int_M \omega(A) + \int_{\partial M} \omega_1(A,g)
\end{equation}
for some local boundary term $\omega_1(A,g)$ which depends on both the gauge background $A$ and the parameter of the gauge transformation $g$. In order to make everything gauge invariant we need to either restrict the gauge transformation near the boundary or introduce new degrees of freedom which transform in an opposite way under gauge transformation.

% Such systems which are invariant under global $G$ transformations but not local ones are said to be anomalous, and we thus obtain a correspondence between ('t Hooft) anomalies and invertible topological gauge theories\cite{}. Ungauging, one leads to usual conclusions about boundaries of SPT phases \cite{CGLW}.

% In the condensed matter literature, a gapped, nondegenerate system with a $G$ symmetry which when gauged produces a Dijkgraaf-Witten theory as above is called an SPT \cite{CGLW}. From what we have just computed, one finds that the boundary of such a system either breaks the symmetry (so when we gauge the gauge transformations there are restricted) or the symmetry is anomalous (meaning there are new degrees of freedom which while $G$-symmetric, transform nontrivially under local $G$ transformations). In \cite{KapustinThorngrenAnomaly1,KapustinThorngrenAnomaly2}, the authors considered the possibility that the $G$ symmetry was projective on the boundary, leading to an extension
% \[H \to \hat G \to G,\]
% where $H$ is a gauge symmetry of the boundary. We will consider analogous boundary conditions for our topological sigma models. For now we note that the important condition is that $\omega$ restricted to the boundary symmetry group (which could be smaller or larger than $G$) becomes exact in cohomology.

The key observation that allows us to port this reasoning to the twisted sigma model case is that gauge transformations of $A$ are equivalent to \emph{homotopies} of the classifying map $A:M \to BG$. Instead of gauge invariance then, we will be interested in how the topological term \eqref{topresp} transforms under homotopies of $\sigma$ when $\partial M \neq 0$. In fact, there is a direct generalization of Eq. \eqref{boundarygaugevariation}. To derive it, we consider a typical homotopy $h:[0,1] \times M \to X$\footnote{Note that a homotopy is a special case of a cobordism.}, where $h(0):M \to X$ is the initial configuration and $h(1)$ the final one. Then, because $\omega$ is closed,
\[0 = \int_{[0,1]\times M} h^*d\omega = \int_{\partial([0,1] \times M)} h^*\omega.\]
Decomposing the boundary of $[0,1] \times M$ into $\{1\} \times M$, $\{0\} \times M$, and $[0,1] \times \partial M$, we then obtain
\[0 = \int_M h(1)^*\omega - \int_M h(0)^*\omega + \int_{[0,1] \times \partial M} h^*\omega.\]
The last term may be written as an integral over $\partial M$ since $[0,1] \times \partial M$ collapses onto $\partial M$. Let this define $\omega_1(h)$. Rearranging, we find a formula analogous to \eqref{boundarygaugevariation}:
\begin{equation}\label{boundaryvariation}
\int_M \sigma'^*\omega = \int_M \sigma^*\omega + \int_{\partial M} \omega_1(h)
\end{equation}
where $h$ is a homotopy from $\sigma$ to $\sigma'$. Therefore, to preserve homotopy invariance, we need to either restrict the homotopies along the boundary (akin to symmetry breaking) or include new degrees of freedom there which transform nontrivially under homotopies (a kind of anomaly).

To be precise, a homotopy invariant boundary condition must have a partition function $Z(\sigma,\partial M)$ which satisfies the anomaly equation
\begin{equation}\label{anomaly}
\delta_h \log Z(\sigma,\partial M) = i\int_{\partial M} \omega_1(h).
\end{equation}
This equation is preserved by renormalization (up to the addition of variations of local terms), since the bulk is topological and does not flow. One way to think about it is we have a boundary theory $Z(\sigma,N)$ for each $\sigma:N \to X$ and a connection $\omega_1$ on the space of these theories, acting as in Eq \eqref{anomaly}. If this connection were flat, then we could use it to canonically identify the $Z(\sigma,N)$ at different $\sigma$ representing the same homotopy class. However, it is not flat. Its curvature is the topological density $\omega$, so these identifications must depend on the path taken, which is exactly what the bulk keeps track of. A similar perspective on anomaly was explained in \cite{cspt}. One can liken it to a fractional Berry connection.

\section{Examples}\label{examples}

% \subsubsection{$m=2$, $X = \mathbb{RP}^2 \times S^1$}

% Consider first a 1+1D system with a map $\sigma:M^2 \to \mathbb{RP}^2 \times S^1$. $\sigma$ could describe, for example, the nematic order parameter in a liquid crystal coupled to .

% There is one possible twist, since $H^2(\mathbb{RP}^2,U(1)) = \bZ_2$. If we write $\alpha$ for the generator of $H^1(\mathbb{RP}^2,\bZ_2)$, dual to $\pi_1 \mathbb{RP}^2 = \bZ_2$, this topological term may be written $\frac{1}{2} \alpha^2$. There is a map from the 2-torus $M = T^2$ to $\mathbb{RP}^2$ where the two cycles of $T^2$ both wrap $\pi_1 \mathbb{RP}^2$.

\subsubsection{$m = 2$, $X = S^2$, bosonic}

For 1+1D theories with a map $\sigma:M^2 \to S^2$, there is a possible $\theta$-term, coupling to the degree of $\sigma$. If we write $\sigma(u,v)$ as a normal vector in $\bR^3$, the $\theta$-term may be written in coordinates $u,v$ on $M$ as
\begin{equation}\label{topterm22}
\frac{\theta}{4\pi} \int_M \sigma \cdot (\partial_u \sigma \times \partial_v \sigma) du dv.
\end{equation}
This corresponds to Eq. \ref{topresp} for $\omega$ the generator of $H^2(S^2,\bZ) = \bZ$. For closed $M$ it is a homotopy invariant of $\sigma$.

We will consider the antipodal map on $S^2$ a symmetry of the problem, however we do not require use of a round metric on the sphere, and indeed there are reasons to expect that round metrics give rise to very special field theories. The only antipodally-symmetric values of $\theta$ are then 0 and $\pi$. This symmetry descends from charge conjugation symmetry of the Abelian Higgs model and so we will denote it $C$ \cite{Komargodskietal}.

% Let $I$ denote the instanton operator in the theory and consider a $C$-symmetric ground state $C|0\rangle = \pm |0\rangle$. We have $C^{\dagger}IC = I^{\dagger}$ and so
% \[\langle 0|I|0\rangle^* = \langle 0|I^\dagger |0\rangle = \langle 0 |C^\dagger I C |0\rangle = \langle 0 | I | 0\rangle\]
% is real. This is also computed by the partition function with $M$ a torus and $[\sigma]$ of degree 1. This indicates that as long as that partition function is nonvanishing it has definite sign. In our terminology of Section I, this means $\theta$ is a well-defined property of a nondegenerate sigma model of this type.

Now suppose $M$ has boundary. The integral above is no longer homotopy invariant. Indeed, we can stretch the boundary over $S^2$ as many times as we like, changing the winding number. In other words, the boundary can continuously absorb or emit instantons. There are several ways of dealing with this. First, if we allow the $C$ symmetry to be broken on the boundary, we may tune $\theta$ continuously to zero in a neighborhood of the boundary and have no problems. If we insist on preserving $C$, however, we will have to impose some boundary condition on $\sigma$.

The simplest boundary condition on $\sigma$ is simply that it is constant along the boundary, say mapping every boundary component to the north pole. Then $M$ may be closed up by gluing to the boundary a set of discs along which $\sigma$ is extended by a constant map.

A more complex boundary condition restricts $\sigma$ to lie along the equator at the boundary. This might come from a planar anisotropy on the boundary. Then we may use a similar trick to define the degree of $\sigma$, again capping off the boundary using discs but now extending $\sigma$ over these discs to that the image lies in the southern hemisphere. The resulting winding number is homotopy invariant.

These are all boundary conditions which in Dijkgraaf-Witten models would be considered symmetry breaking boundary conditions, but there are also truly anomalous boundaries. An interesting one in the case at hand (which is in some sense minimal) allows $\sigma$ to be unconstrained at the boundary and in fact adds a circular degree of freedom $\rho:\partial M \to S^1$. The combined target space of $(\rho,\sigma)$ is taken not to be the product but rather a non-trivial fibration: the Hopf fibration
\[S^1 \xrightarrow{i} S^3 \xrightarrow{\pi} S^2.\]
So that the boundary degrees of freedom $(\rho,\sigma)$ can be combined into a single map $\hat \sigma:\partial M \to S^3.$ This is analogous to considering a non-trivial extension of boundary gauge groups, which was used to construct anomalies in \cite{KapustinThorngrenAnomaly, WenWitten}.

The key fact that we use about the Hopf fibration is that the winding number around the fiber is not well defined. To see why, we begin by considering a vector field $\vec w$ on $S^3$ which points along the fibers and along any fixed one $S^1$ is constant and the line integral $\int_{S^1} \vec w \cdot \hat t dt = 1$. The fibers twist around each other, so necessarily this vector field will have some non-trivial dependence on the $S^2$ coordinate $\sigma$ which parametrizes the fibers. In fact, the divergence of $\vec w$ is the pullback of the volume form on $S^2$. It follows that the combined bulk-boundary topological term
\begin{equation}\label{bulkplusboundary}
\frac{\theta}{4\pi} \int_M \sigma \cdot (\partial_u \sigma \times \partial_v \sigma) du dv + \theta \int_{\partial M} \hat \sigma^*w
\end{equation}
is homotopy invariant. Note that C acts by reversing the orientation along the fiber.

The boundary sigma model cannot exist as a purely 0+1D system since it would be a particle that sees a half quantum of magnetic flux through the Hopf fiber, by \eqref{bulkplusboundary}. Compare with the anomaly in Appendix D of \cite{GKKS}. Equivalently, if one takes to be $\sigma:\partial M \to S^2$ as a slowly varying background parameter for the boundary theory, the boundary anomaly amounts to a Berry curvature of $\pi$ over $S^2$. String theorists should compare with D branes in presence of $B$ field \cite{KapustinDBranes}.

One can derive the same conclusions for the boundary of the dynamical model from the $N=2$ Abelian Higgs model, by considering only gauge transformations which are the identity along the boundary. This frees a circular degree of freedom leaving semiclassical Higgs moduli space $\{|\phi_1|^2 + |\phi_2|^2 = 1\} = S^3$.

Note that when $\theta = 2\pi$, the $\theta$ angle can be written as a boundary "WZW" term and no new degrees of freedom are needed\cite{XuLudwig,Bietal}. In this case, we have ordinary bulk-boundary SPT physics.

\subsubsection{$m = 2$, $X = S^1$, fermionic}

It is also possible to give interesting examples in fermionic systems. For example, we can consider a 1+1D system depending on a map $\sigma:M^2 \to S^1$ as well as a spin structure $\eta$ on $M$. There is a single twist in spin cobordism $\Omega^2_{spin}(S^1) = H^1(S^1,\Omega^1_{spin}(\star)) = H^1(S^1,\bZ_2) = \bZ_2$. One way to express this topological term is to obtain from $\eta$ the associated quadratic form $q_\eta:H^1(M,\bZ_2) \to \bZ_2$ \cite{KirbyTaylor,KTTW}. Then, writing $\alpha$ as the generator of $H^1(S^1,\bZ_2) = \bZ_2$, we obtain a sign $\exp i\pi q_\eta(\sigma^*\alpha).$

The bulk meaning of this term is that when the string wraps the target circle an odd number of times, the ground state has odd fermion parity. When the string has boundary, the winding number becomes ill-defined and so does the fermion parity \cite{Baezetal}. However, it is possible to make the winding number defined mod 2 if we lift the boundary map to the double cover circle $S^1 \xrightarrow{2} S^1$. This is an extra $\bZ_2$ degree of freedom at the boundary: a fermionic zero mode.

We may also consider the boundary theory as a family of quantum mechanical theories parametrized by $S^1$. Such a family has a Berry phase and from the above consideration of the fermion parity, we see the meaning of the anomaly is that the holonomy of the Berry connection exchanges states of opposite fermion parity. This guarantees the robustness of the zero mode.

\subsubsection{$m = 3$, $X = S^2$, fermionic}\label{fermionicskyrmion}

Our next example is obtained from abelian gauge theory by Higgs mechanism. We begin with a $U(1)$ gauge field $a$ with Chern-Simons term at level 1 coupled to two charge 1 scalars $\phi_1,\phi_2$. Then we turn on an $SU(2)$-symmetric Higgs potential. As analyzed in \cite{PolyakovWiegmann,FKS}, this results in a sigma model of maps $\sigma:M^3 \to S^2$ with a special topological term for which the Skyrmion has odd fermion parity. This topological term lives in $H^2(S^2,\Omega^1_{spin}(\star)) = H^2(S^2,\bZ_2) = \bZ_2$. When $M = S^3$ with its unique spin structure, this topological term is minus one to the Hopf number of $\sigma:S^3 \to S^2$.

What happens when we allow $M$ to have boundary? In the gauge theory, we need to deal with the boundary variation of the Chern-Simons term. This is done by adding a chiral scalar $\rho:\partial M \to S^1$ to the boundary \cite{Wen,Elitzuretal}. The total charge of this chiral scalar is equal to the winding number of this map \cite{Tong}. In terms of the 1-form $w$ of the previous example, this is $\int_{\partial M} \rho^*w$. However, because of the chiral anomaly, this charge is not conserved in the presence of $a$ instantons \cite{Adler,BellJackiw}. Instead, $d\rho^*w = da/2\pi$.

When we go to the Higgs phase, described by the sigma model, the chiral anomaly equation becomes $d\rho^*w = \sigma^*\omega$, reproducing the relation derived between fiber and base winding numbers for the Hopf fibration above. The conclusion is that along the boundary, the $S^2$ residual Higgs degree of freedom combines with the chiral scalar to produce map $\hat \sigma:\partial M \to S^3$. Then, for fermion parity to be well-defined, the winding number $\rho^*w$ around the Hopf fiber must also contribute to the fermion parity, and only the combination of the two is conserved.

% \subsubsection{$m = 4$, $n = 2$, fermionic}

% In 3+1D, we can consider a twist coming from $H^2(X,\Omega^2_{spin}) = H^2(X,\bZ_2)$. For example, with $X=S^2$, this term gives the ``Hopfion" $S^3 \to S^2$ ($\sigma$ in the homotopy class of the Hopf fibration) odd fermion parity. ...

% its Euler class is the generator $\omega \in H^2(S^2,\bZ)$ \cite{}. This means that if $\alpha$ is the volume form of $S^1$, normalized to $\int_{S^1} \alpha = 1$ and extended to $S^3$ in any particular fashion by $i_* \alpha$, it will satisfy
% \begin{equation}\label{hopfanomaly}
% di_*\alpha = \pi^* \omega.
% \end{equation}

% If we consider $M$ with boundary, then to preserve homotopy invariance we must either break the antipodal symmetry or change the target space at the boundary. For example, let's restrict $\sigma|_{\partial M}$ to live in the equator $S^1$ and focus on a single boundary component, so that it becomes a map $\sigma|_{\partial M}:S^1 \to S^1$. It is always possible to extend such a map to the disc $\sigma':D^2 \to D^2$. We consider the target as the southern hemisphere of the target sphere and use this extension to extend $\sigma$ to the \emph{closed} surface $\hat \sigma:M \cup_{\partial M} D^2 \to S^2$. It is easy to show that the winding number of $\hat \sigma$ only depends on the homotopy class of the constrained $\sigma$. What we have done is analogous to adding WZW terms to the 

\section{topological transitions}\label{transition}

Let us return to our assumption of nondegeneracy. When this assumption is violated by the closing of a gap in some sector $(N,[\sigma])$ we may not be able to associate a topological term to the special point \`a la Eq. \eqref{gap}. Therefore, conclusions about protected boundary behavior must be discarded and we may have the analog of a bulk transition in SPT phase \cite{Lokmanetal}.

To have an example, consider the abelian Higgs model setup of example \ref{fermionicskyrmion} coupled as well to a charge 1 Dirac fermion $\psi$ with mass $m_{\psi}$. We regulate the fermion determinant so that at positive mass $m_{\psi}>0$ no Chern-Simons term is generated upon integrating $\psi$ while for $m_{\psi}<0$ a level 1 Chern-Simons term is generated by parity anomaly \cite{Redlich}. Meanwhile we consider an $SU(2)$-symmetric Higgs potential for the scalars. In the deep Higgs phase, with $m_{\psi}\gg 0$, we obtain from $(\phi_1,\phi_2)$ an untwisted sigma model into $S^2$. While at $m_{\psi} \ll 0$, the Chern-Simons term leads to the twisted sigma model into $S^2$ we studied in example \ref{fermionicskyrmion}. Along the boundary, in the second case we have an $S^3$ sigma model which splits after the transition into an $S^2 \times S^1$ sigma model, where the $S^1$ is no longer protected by the chiral anomaly and may be gapped out.

In the bulk, the two sigma models are distinguished by the fermion parity of the skyrmion. However, at $m_{\psi} = 0$, the Dirac fermion has a zero mode in the presence of a skyrmion. Therefore, at this special point, there are two skyrmion ground states with opposite fermion parity. In particular, according to our prescription, we cannot associate an invertible topological sigma model to this special point. This is because $Z(S^2 \times S^1,\sigma,\eta)$ with $\sigma$ wrapping $S^2 \to S^2$ once and the non-bounding, periodic spin structure $\eta$ around the circle, computes $\Tr{(-1)^F \exp -\beta H}$ in this sector, hence $Z(S^2 \times S^1,\sigma,\eta) = 0$ because of the zero mode.

From the perspective of the infrared, at the $m_\psi = 0$ point, a new gapless degree of freedom enters. It would be interesting to understand whether this is typical of such ``topological" transitions and which pairs of twisted sigma models admit generically direct transitions.

\section{Acknowledgements}

I am pleased to acknowledge D. Else, A. Kapustin, Z. Komargodski, and M. Metlitski for enlightening discussions and related collaborations as well as funding from the NSF GRFP.

% \bibliography{references}


\end{document}